\RequirePackage{fix-cm}
\documentclass[smallextended,english]{svjour3}       
\smartqed  
\usepackage{graphicx}
\usepackage[T1]{fontenc}
\usepackage[latin9]{inputenc}
\usepackage{amsmath}
\usepackage{graphicx}
\usepackage{amssymb}
\usepackage{esint}
\usepackage{epstopdf}
\DeclareGraphicsExtensions{.pdf,.eps,.png,.jpg,.mps} 

\makeatletter
\@ifundefined{textcolor}{}
{%
 \definecolor{BLACK}{gray}{0}
 \definecolor{WHITE}{gray}{1}
 \definecolor{RED}{rgb}{1,0,0}
 \definecolor{GREEN}{rgb}{0,1,0}
 \definecolor{BLUE}{rgb}{0,0,1}
 \definecolor{CYAN}{cmyk}{1,0,0,0}
 \definecolor{MAGENTA}{cmyk}{0,1,0,0}
 \definecolor{YELLOW}{cmyk}{0,0,1,0}
 }

\newcommand{\be}{\begin{eqnarray}}\newcommand{\ee}{\end{eqnarray}}\def\beq{\begin{equation}}\def\eeq{\end{equation}}

\makeatother

\usepackage{babel}

\makeatother

\begin{document}
\title{Inhomogenous pairing and enhancement of superconductivity in large Sn nanograins}
\author{James Mayoh and Antonio M. Garc\'{\i}a-Garc\'{\i}a}
\institute{
              TCM group, Cavendish Laboratory, JJ Thomson Avenue, Cambridge, CB3 0HE, UK.}
\date{Received: date / Accepted: date}
\maketitle
\begin{abstract}
A substantial enhancement of the superconducting gap was recently reported in clean, large $\sim 30$nm, and close to hemispherical Sn grains. A satisfactory explanation of this behaviour is still missing as shell effects caused by fluctuations of the spectral density or surface phonons are negligible in this region. Here we show that this enhancement is caused by spatial inhomogeneities of the Cooper's pairs density of probability. In the mean field approach that we employ these inhomogeneities are closely related to the eigenstates of the one-body problem, namely, a particle in a hemispherical shaped potential. The parameter free theoretical prediction agrees well with the experimental results. A similar enhancement is predicted for other weakly coupled superconductors.
\end{abstract}
\section{Introduction}
Experimental reports, starting in the sixties \cite{abeles}, of substantial enhancement of superconductivity in thin granular films of different materials have been a continuous stimulus to study superconductivity in low dimensions. However the dramatic increase of the critical temperature observed in materials like Al or Sn \cite{abeles} resisted a conclusive theoretical explanation. The cause of the enhancement was related to surface phonons, fluctuations of the spectral density around the Fermi energy or shape resonances \cite{blat,bianconi}. The first two proposals could not be reconciled with the fact that the enhancement was observed on some materials but not in others. The latter mechanism, put forward by Blatt and Thompson \cite{blat}, is only effective for clean thin films with only a few mono-layers thick. However the samples were granular and disordered. Indeed more refined experimental studies \cite{goldman,qsize} where thin films were smoother and granularity was attenuated showed no substantial enhancement of superconductivity. 
In the context of single nanograins, the seminal experiments of Tinkham and coworkers \cite{tinkham} on single nanoscale Al grains provided evidence that some sort of superconductivity is still present in grains of only a few nanometer size where the mean level spacing is comparable with the energy gap \cite{ander2}. 
For larger clean grains, but still within the nanoscale region, numerical solutions of the the Bardeen-Cooper-Schrieber (BCS) gap equation \cite{bcs} and Boguliubov de Gennes equations \cite{parmenter,fomin,shell,shanenko,peeters,heiselberg} showed that the critical temperature and other superconducting properties were highly non-monotonic as a function of the system size with peaks well above the bulk limit. Explicit results were obtained for a variety of shapes and confining potentials:cubes \cite{parmenter,fomin}, spheres \cite{shell,shanenko}, cylinders \cite{peeters} and harmonic confining potentials \cite{heiselberg}.  The magnitude of the peaks, namely, the enhancement of superconductivity, was larger in spherical and cubic grains than in chaotic grains \cite{usprl,leboeuf} with no symmetry. Moreover, for a fixed size, deviations from the bulk limit are more pronounced as the superconducting coherence length of the material increases. Analytical results \cite{usprl,leboeuf} based on the periodic orbit theory \cite{baduri} indicate that the reason for these non-monotonic deviations from the bulk limit was associated to shell effects, namely, level degeneracy in the proximities of the Fermi energy due to the geometrical symmetries of the grain. A larger spectral density induces an effective stronger binding of Cooper's pairs that boost superconductivity. Recent experiments on single isolated hemispherical Sn grains \cite{nmat} have reported, in full agreement with the theoretical prediction \cite{usprl,nmat}, large oscillations in the size dependence of the energy gap in the region $\sim 10$nm. However some puzzles still remain. For instance, in \cite{nmat} it was also observed a substantial ($\approx 20\%$) monotonic enhancement of the superconducting gap up to the largest grains $\approx 30$nm studied that cannot be explained by shell effects or surface phonons. 
\begin{figure}
\includegraphics[width=1\columnwidth,clip,angle=0]{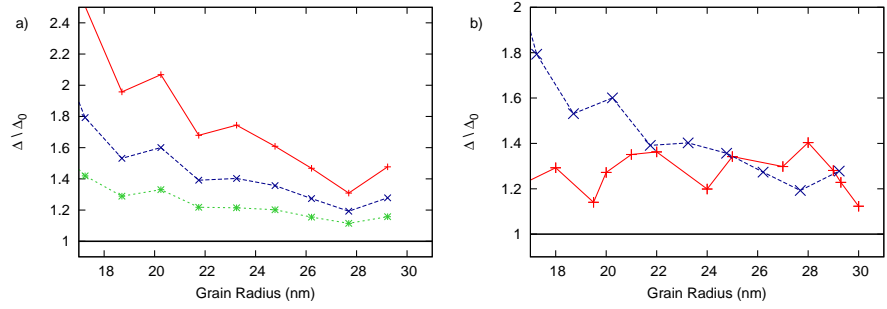}\caption{\label{fig1}(a) The mean superconducting gap as a function of the hemispherical grain size for $\lambda=0.166$ Al (dotted line),$\lambda = 0.243$ Sn (dashed line), $\lambda = 0.382$ Pb (solid line). (b) Comparison between the experimental results (solid line) of Ref. \cite{nmat} for Sn hemispherical nanograins and the theoretical prediction (dashed line) Eq.(\ref{gapsup}) that includes the effect of inhomogeneous pairing. We have averaged fluctuations in order to single out the contribution not related to shell effects in the spectral density. The horizontal line corresponds to the bulk behaviour.}
\end{figure}

Here we provide evidence that this monotonic enhancement is caused by spatial fluctuations in the density of probability of Cooper's pairs in a confined geometry. We carry out a numerical calculation, within a mean-field framework, of the order parameter in a hemispherical grain for sizes up to $30$nm. Our results, see Fig. \ref{fig1}, which are parameter free, are in fair agreement with recent experimental results \cite{nmat}. This additional enhancement stems from the fact that, in finite size grains, the interactions that bind the electron into a Cooper's pair depend on the quantum numbers of the one-body problem eigenstates. The dimensionless electron-phonon coupling constant $\lambda$ becomes inhomogeneous as it depends on these quantum numbers, $\lambda \to \lambda V I_{n,m}$ where $V$ is the grain volume, $I_{n,m} = \int dr^d \Psi^2_n(r)\Psi^2_m(r)$
and $\Psi_n(r)$ is the eigenstate of the one body problem and $n$ the set of quantum numbers that labels the state. For the case of grains with no symmetry, the leading finite size correction due to this effect Ref. \cite{vinas} is positive $I = 1+A/k_FL$ with $A \geq 0$ that depends on boundary conditions. For a chaotic grain the semiclassical analytical analysis of Ref. \cite{usprl} showed that this the leading correction for sizes $L \geqslant 10$nm. 

 \section{The model and results}
The superconducting grain is described by the BCS Hamiltonian \cite{bcs},
\begin{equation}
H=\sum_{{\bf n}\,\sigma}\epsilon_{\bf n} c^\dag_{{\bf n}\sigma}c_{{\bf n}\sigma}-\frac{\lambda}{\nu(0)}\sum_{{\bf n},{\bf n'}}I_{{\bf n},{\bf n'}}c_{{\bf n}\uparrow}^\dag c_{{\bf n}\downarrow}^\dag c_{{\bf n'}\uparrow}c_{{\bf n'}\downarrow}
\end{equation}
where $c_{{\bf n}\sigma}^\dag$ creates an electron of spin $\sigma$ in a state with quantum numbers ${\bf n}$ and energy $\epsilon_{\bf n}$, $\lambda$ is the dimensionless BCS coupling constant for the material, $\nu(0)$ is the density of states at the Fermi energy. The electron-electron interaction matrix elements resulting from a contact interaction is given by,
\begin{equation}
I_{{\bf n},{\bf n'}}=V\int \psi_{\bf{n}}^2({\bf r})\psi_{\bf{n'}}^2({\bf r})\,dV
\label{Mel}
\end{equation}
where $V$ is the volume of the grain and $\psi_{\bf{n}}({\bf r})$ is single-electron eigenfunction in state $\bf{n}$.\\
The superconducting gap is calculated from the self-consistency equation,
\begin{equation}
\Delta_{\bf n}=\frac{\lambda}{2}\displaystyle\sum_{{\bf n'}}\frac{\Delta_{{\bf n'}}I_{{\bf n},{\bf n'}}}{\sqrt{\epsilon_{{\bf n'}}^2+\Delta_{{\bf n'}}^2}}\frac{1}{\nu(0)}
\label{GapEqn}
\end{equation}
where the sum is now taken over all states $\left\{{\bf n'}\big|\:|\epsilon_{{\bf n'}}|<\epsilon_D \right\}$, and $\epsilon_D$ is the Debye energy. We note that \cite{shanenko} that this approach leads to results similar to those obtained from the technically more involved Bogoliubov de Gennes equations. 
In the bulk limit eigenfunctions are plane waves and the matrix elements are simply  $I_{{\bf n},{\bf n'}}=1$ 
 However in small grains eigenstates of the one body problem are not plane waves so we expect deviations in $I_{n,n'}$ from the bulk limit. 
We restrict our interest to grains such that $\delta/\Delta_0 \ll 1$
where a BCS mean field approach is valid.
To make direct comparison with experimental results we calculate numerically $I_{n,n'}$ for hemispherical grains of radius $R$. We note that the experimental grains \cite{nmat} are not exactly hemispherical but closer to a spherical cap of height $h \sim 0.9R$. Although the non-monotonic oscillations due to shell effects depend strongly on the shape of the grains we expect that the monotonic deviations that we aim to describe are less sensitive to this relatively small shape difference.  Therefore the eigenfunctions entering in the matrix element above are those of a single electron in a spherical grain of radius $R$,
\begin{equation}\label{hemiWF}
\psi_{n,l,m}(r,\theta,\phi)=Nj_l(u_{ln}\frac{r}{R})Y_{lm}(\theta,\phi)
\end{equation}
where $N=\frac{2}{j_{a+1}(u_{am})R^{3/2}}$ is the normalisation constant, $j_l(r)$ are the spherical Bessel functions of the first kind, $u_{ln}$ is the $n^{th}$ zero of the $l^{th}$ spherical Bessel function and $Y_{lm}(\theta,\phi)$ are the spherical harmonics. The energy associated with these eigenstates is, 
$E_{l,n}=\frac{\hbar^2u_{ln}^2}{2mR^2}$. Dirichlet boundary conditions on the hemispherical surface \cite{Rodriguez2001} restricts $|m-l|$ to be odd. 

The final expression for the matrix elements is simplified by using Clebsch-Gordan coefficients,
\begin{eqnarray}
I_{{\bf n},{\bf n'}} = \frac{4(2l+1)(2l'+1)}{3 j_{a+1}(u_{am})^2j_{a'+1}(u_{a'm'})^2}\sum_{\Lambda}\frac{<ll',mm'|ll',\Lambda M>^2<ll',00|ll',\Lambda0>^2}{(2\Lambda+1)} \nonumber \\ \nonumber \times \int_0^1j_l(u_{ln}\rho)^2j_{l'}(u_{l'n'}\rho)^2\rho^2d\;\rho
\end{eqnarray}
where $M=m+m'$ and $\Lambda$ is summed over all possible values in the range, $l+l'\geq \Lambda$,
$|l-l'|\leq\Lambda$ and $m\leq|\Lambda|$. 

The superconducting gap can then be written as, 
\begin{equation}
\Delta=2\epsilon_D e^{-\frac{1}{\lambda_{eff}}}
\label{gapsup}
\end{equation}
where $\lambda_{eff}=\lambda {\bar I}$ and $\bar I$ is the average of $I_{{\bf n},{\bf n'}}$ over all states in the interacting region $2\epsilon_D$ where $n'$ is the level closest to the Fermi-energy. This should be a good approximation for sufficiently large grains for which the matrix elements do not depend strongly on the quantum numbers. This is also consistent with the observation that in scanning tunnelling  microscope experiments \cite{nmat} the value of the gap did not depend much on the exact position of the tip. Moreover it was found in \cite{usprl} that a similar simplified expression for the gap describes shell effects related to fluctuations of the spectral density. In that case the resulting spectral density after solving the gap equation is expressed as a finite sum over classical periodic orbits of length less than the superconducting coherence length.

 The numerical results are shown in Fig. \ref{fig1} for $\lambda=0.243$. This value is consistent with the Sn bulk gap $\Delta_{bulk}=0.57$meV and a Debye energy $\epsilon_D=17.2$meV. The numerical results, see Fig \ref{fig1}, show substantial deviations from the bulk even at large grain sizes. The theoretical prediction is strikingly similar to the experimental observation. We stress that the theoretical expression is parameter fee. Results for other materials, see Fig. \ref{fig1}, are similar. From Eq. (\ref{gapsup}) it is clear that finite size effects are stronger the smaller is the coupling constant $\lambda$. In more physical terms, finite size effects are stronger in materials with a long superconducting coherence length $\xi \propto 1/\Delta \propto e^{1/\lambda}$. A few final comments are in order: a) shell effects that induce oscillations in the spectral density have been averaged out in order to single out monotonic deviations from the bulk limit, b) the dip at $\sim 28$nm in the theoretical prediction is likely a consequence of statistical fluctuations related to the relatively small number of points employed in the averaging of gap size oscillations, c) deviations for smaller sizes $ < 18$nm are likely due to the difference between the spherical cap shape of the experimental grains and the exact hemispherical shape employed in the theoretical calculation. 

In conclusion we have investigated superconductivity in hemispherical nano-grains of metallic superconductors. Deviations from the bulk are clearly observed even for the largest grains $\sim 30$nm. Experimental results \cite{nmat} in single isolated Sn nanograins are in full agreement with the analytical predictions. Similar results are expected in other weakly coupled superconducting materials. 

\end{document}